\documentclass{article}

\usepackage{epsfig}

\begin{document}

\title{Experimental signature of a fermiophobic Higgs boson}



\author{ {L. BR\"UCHER \footnote{e-mail: bruecher@alf1.cii.fc.ul.pt},
R. SANTOS \footnote{e-mail: rsantos@alf1.cii.fc.ul.pt}}\\
Centro de F\'\i sica Nuclear da Universidade de Lisboa,\\
Av. Prof. Gama Pinto 2, 1649-003 Lisboa, Portugal}


\maketitle

\abstract{
  The most general Two Higgs Doublet Model potential without explicit
  $CP$ violation depends on 10 real independent parameters . Excluding
  spontaneous $CP$ violation results in two 7 parameter models.
  Although both models give rise to 5 scalar particles and 2 mixing
  angles, the resulting phenomenology of the scalar sectors is
  different.
  If flavour changing neutral currents at tree level are to be
  avoided, one has four alternative ways of introducing the fermion
  couplings in both cases. In one of these models the mixing angle of
  the $CP$ even sector can be chosen in such a way that the fermion
  couplings to the lightest scalar Higgs boson vanishes. At the same time it is
  possible to suppress the fermion couplings to the charged and
  pseudo-scalar Higgs boson by appropriately choosing the mixing angle of
  the $CP$ odd sector.
  We investigate the phenomenology of both models in the fermiophobic
  limit and present the different branching ratios for the decays of
  the scalar particles. We use the present experimental results from
  the LEP collider to constrain the models.
}

\section{Introduction}

Despite the great success of the Standard model (SM) the mechanism to
generate the vector boson masses, the so called Higgs mechanism, still
awaits experimental confirmation. Current limits at LEP yield a mass
of $m_h > 91.0\ GeV$ \cite{SMH} for a minnimal Higgs boson. Thus it is
appropriate to investigate models with an extended Higgs sector, which
allow a light Higgs boson not restricted by the current SM Higgs mass
limit. A class of these models are the Two Higgs Doublets
models (2HDM) with type I coupling to the fermions \cite{Sant1}. In the
following we will discuss these models in the so-called fermiophobic
limit. We start our discussion with summarizing the 2HDM potentials and
defining the fermiophobic limit. Thereafter we will restrict the
physical parameters by theoretical constraints. Then we will discuss
the branching ratios of the light scalar Higgs particle. Finally we
will constrain the model by using recent experimental data.

\section{The potentials}

The most general 2HDM potential invariant under $SU(2)\times U(1)$ has
fourteen independent real parameters. If one imposes that the
potential neither explicit nor spontaneously violates $CP$ one has two
different possibilities to restrict the potential \cite{Sant3}. First, the
potential can be made invariant under a $Z_2$ transformation $\phi_1
\rightarrow \phi_1$ and $\phi_2 \rightarrow -\phi_2$. The resulting
potential, which is known as $V_A$, is:
\begin{equation}
V_{A}=-\mu_{1}^{2} x_{1}-\mu_{2}^{2} x_{2}+\lambda_{1} x_{1}^2
 +\lambda_{2} x_{2}^2+\lambda_{3} x_{3}^2+\lambda_{4} x_{4}^2 
 +\lambda_{5} x_{1} x_{2} \enskip , 
\end{equation}
where we used the abbriviations $x_1 = \phi_1^\dagger \phi_1$, $x_2 =
\phi_2^\dagger \phi_2$, $x_3 = \Re\{\phi_1^\dagger \phi_2\}$ and $x_4
= \Im\{\phi_1^\dagger \phi_2\}$.
Second, it is possible to make the potential invariant under the
global $U(1)$ transformation $\phi_2 \rightarrow e^{i \theta} \phi_2$.
The potential then reads:
\begin{equation}
V_{B} =-\mu_{1}^{2} x_{1}-\mu_{2}^{2} x_{2} -\mu_{12}^{2} x_{3}
 +\lambda_{1} x_{1}^2+\lambda_{2} x_{2}^2
 +\lambda_{3} \left( x_{3}^2+ x_{4}^2 \right) 
 +\lambda_{5} x_{1} x_{2} \enskip . 
\end{equation}
Note that the term $-\mu_{12}^{2} x_{3}$ breaks the global symmetry
softly. Both $V_A$ and $V_B$ have seven
degrees of freedom, the four particle masses, the two rotation angles
($\alpha, \beta$) and the term providing the masses for the gauge
bosons. The major difference of the potentials is in the scalar self
couplings. This leads to a different phenomenology not only in the
cases where the Higgs particles interact among themselves, but also
when loop effect play a dominant role in particle decays.

\section{The fermiophobic limit}\label{masslim}

\begin{figure}[htbp]
  \begin{center}
    \epsfig{file=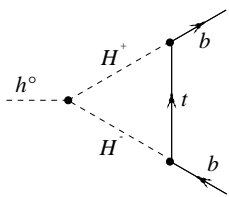,width=3cm}
    \caption{Feynman diagram of the largest contribution to $h^0\rightarrow b \bar{b}$}
    \label{fig:fgs}
  \end{center}
\end{figure}

Although potential $V_A$ and $V_B$ give rise to different scalar
self-couplings, the couplings of the scalars to the fermions and the
vector bosons are the same. Avoiding flavour changing neutral currents
induced by Higgs exchange one has four different ways to couple the
fermions to the Higgs sector. This is done most naturally by extending
the global symmetry to the Yukawa Lagrangian. The resulting for
different models are usually denoted as as model I, II, III and IV
(cf.  e.g. \cite{Sant1}).

In model I all fermions couple to just one Higgs doublet. Thus, by
choosing $\alpha =\pi/2$, one obtains a complete fermiophobic light
$CP$-even scalar Higgs particle, $h^0$, in this model. However,
$h^0$ can still decay to a fermion pair via $h^0 \rightarrow
  W^{*}W (Z^{*}Z) \rightarrow 2\, \bar{f} f$ or $h^0
  \rightarrow W^* W^* (Z^* Z^*) \rightarrow 2\,\bar{f} f$. We will
include these decays in our analysis. Moreover, decays of
$h^0$ to two fermions can also be induced by scalar and gauge boson
loops (see e.g. fig.~\ref{fig:fgs}). But fortunately it turns out that
the only relevant one-loop decay is $h^0
\rightarrow b\bar{b}$ due to a large contribution of the Feynman
diagram shown in fig.~\ref{fig:fgs} to the total decay
width.\footnote{The coupling $[H^+\bar{t}b]$ is proportional to the
  $t$-quark mass.}  Thus, on one hand, $h^0$ is not completely
fermiophobic at $\alpha=\pi/2$, and on the other hand, all decays $h^0
\rightarrow f\bar{f}$ but $h^0 \rightarrow b\bar{b}$ are almost zero
even at one-loop level.
The coupling of the $h^0$ to the vector bosons is proportional to the
sine of $\delta\equiv \alpha-\beta$. If we let $\beta$ tend to $\alpha$
( $\beta \rightarrow \alpha=\pi/2$ ), then $h^0$ is not only
fermiophobic but also bosophobic and ``ghostphobic'' -- It always
needs another scalar particle to be able to decay. The differences
between potential $A$ and $B$ can be extremely important in this limit
since $h^0$ will have different signatures in each model. In contrast,
the heaviest $CP$-even scalar, $H^0$, acquires the Higgs standard
model couplings to the fermions in this limit. We will relax the limit
$\beta \approx \pi/2$ and analyze the decays as a function of $\delta$
and of the Higgs masses.

\begin{figure}[htbp]
  \begin{center}
    \epsfig{file=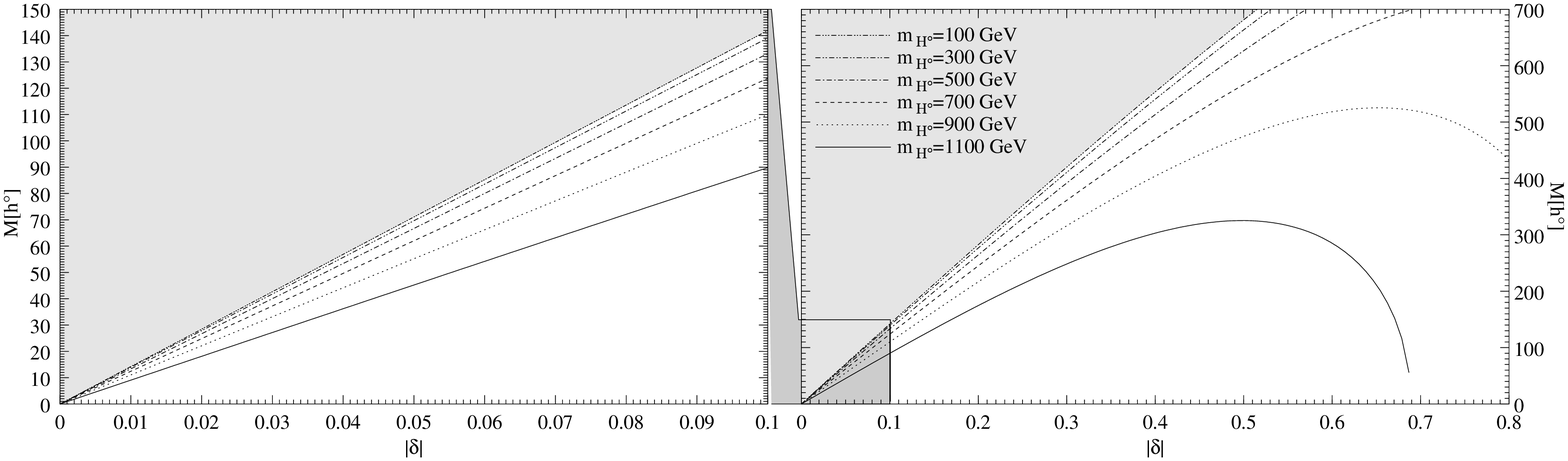,width=8.5cm,height=4.5cm}
    \caption{Limit on $m_{h^0}$ as a function of $\delta$ in potential$A$.}
    \label{fig:Mhlimit1A}
  \end{center}
\end{figure}

\begin{figure}[htbp]
  \begin{center}
    \epsfig{file=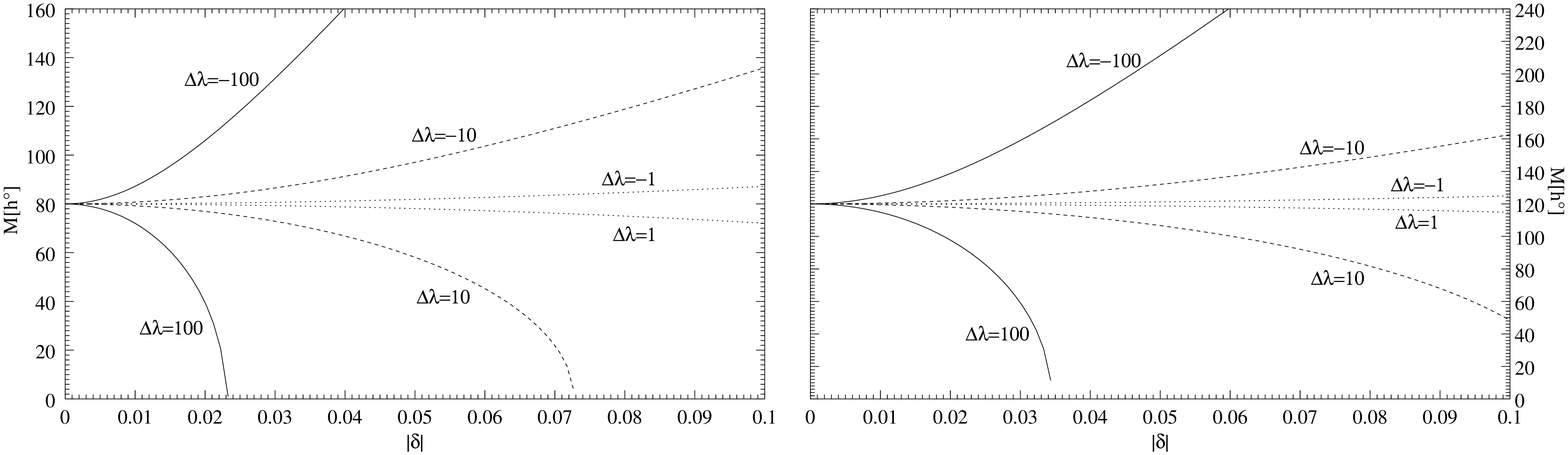,width=8.5cm,height=4.5cm}
    \caption{Limit $m_{h^0}$ as a function of $\delta$ for
      $m_A=80\ GeV$ and $m_A=120\ GeV$.}
    \label{fig:Mhlimit1B}
  \end{center}
\end{figure}

Before we start our analysis we have to ensure, that by choosing a set
of values for $(m_{h^0}, m_{H^0}, m_A, m_{H^+}, \alpha, \beta)$ we do
not leave the perturbative regime. In general, the bounds ensuring
this, are the so-called tree-level unitarity bounds
\cite{trunal1}. For potential $V_A$ they yield:
\begin{equation}
  \label{eq:vacA}
  m_{h^0} \, \le \, \sqrt{\frac{16\pi\sqrt{2}}{3\,G_F}\,\cos^2\beta \,
  - \, m_{H^0}^2 \cot^2\beta} \enskip ,
\end{equation}
where $G_F=1.166\ GeV^{-2}$ denotes Fermi´s constant. We have plotted
this equation in fig.~\ref{fig:Mhlimit1A}. One easily verifies that in
the limit $\delta \rightarrow 0$ $h^0$ becomes massless, which is also
clear from eq.~\ref{eq:vacA}. Unfortunately no tree-level
unitarity bounds are avalaible for potential $V_B$.
Nevertheless, we know that in the fermiophobic limit \cite{Sant4}:
\begin{equation}
  \label{eq:vacB}
   m_{h^0}^2  =   m_A^2 - 2\left(\lambda_+ - \lambda_1\right) v^2
   \cos^2 \beta \enskip ,
\end{equation}
with $\lambda_+ = \frac{1}{2}(\lambda_3+\lambda_5)$ and $v=246\ GeV/c^2$
denoting the vacuum expectation value. The equation shows, that in the
limit $\delta\rightarrow 0$ the masses of $h^0$ and $A^0$ will be
degenerated, which is also illustrated in fig.~\ref{fig:Mhlimit1B}.

The overall picture given by all branching ratios led us to
distinguish between three different regions for $\delta$. We define
these regions now for the following qualitative analysis:
\begin{itemize}
\item the {\em tiny} $\delta$ region where $|\delta| \le 0.05$,
\item the {\em small} $\delta$ region with $0.05 < |\delta| \le 0.1$
  and
\item finally the medium and {\em large} $\delta$ region when $|\delta| > 0.1$.
\end{itemize}

\section{The lightest scalar Higgs boson}\label{h0}

As already pointed out, the lightest scalar Higgs boson ($h^0$) has no
tree level couplings to the fermions for $\alpha=\pi/2$. Thus the
following tree level decays have to be considered:
\begin{eqnarray*}
 && h^0 \rightarrow W^+ W^- \quad ; \quad  
  h^0 \rightarrow Z Z \quad ; \quad 
  h^0 \rightarrow Z A^0 \quad ; \quad \\ &&
  h^0 \rightarrow W^\pm H^\mp \quad ; \quad 
  h^0 \rightarrow A^0 A^0 \quad ; \quad  
  h^0 \rightarrow H^+ H^- \quad .
\end{eqnarray*}
Additionally the following one-loop induced decays are important:
\begin{displaymath}
  h^0 \rightarrow \gamma\gamma \quad ; \quad 
  h^0 \rightarrow Z \gamma \quad ; \quad 
  h^0 \rightarrow b \bar{b} \quad .
\end{displaymath}
Moreover, decays to fermions via virtual vector bosons have to be
taken into account, namely:
\begin{eqnarray*}
  && h^0 \rightarrow W^* W^* \rightarrow f\bar{f} f\bar{f} \quad ; \quad
   h^0 \rightarrow W^* W \rightarrow f\bar{f} W \quad ; \quad \\ &&
   h^0 \rightarrow Z^* Z^* \rightarrow f\bar{f} f\bar{f} \quad ; \quad
   h^0 \rightarrow Z^* Z \rightarrow f\bar{f} Z  \quad .
\end{eqnarray*}
The partial tree-level decay widths are listed in \cite{Bru2}, where
also results for the other Higgs particles $A$, $H^\pm$ and $H^0$
can be found. The one-loop induced decays have been calculated with
{\em xloops} \cite{xl1}. Decays via virtual particles have been calculated in
ref.~\cite{andr}. We have taken these formulas and changed them
appropriately.

As stated earlier, the only significant decay mode to fermions, via
vector boson and scalar loops, is $h^0 \rightarrow b\bar{b}$. For all
the 
other fermionic decays the Feynman graphs are suppressed either by the
Cabbibo-Kobayashi-Maskawa matrix or by the small mass of the fermions
in the loop. However, the diagram shown in
fig.~\ref{fig:fgs} is suppressed by a $\tan^2 \delta$ factor when
compared with the corresponding diagram in $h^0 \rightarrow
\gamma\gamma$.  Thus, as will be seen below, the decay $h^0\rightarrow
b\bar{b}$ is of minor importance in the tiny and small $\delta$
region.

\begin{figure}[htbp]
  \begin{center}
    \epsfig{file=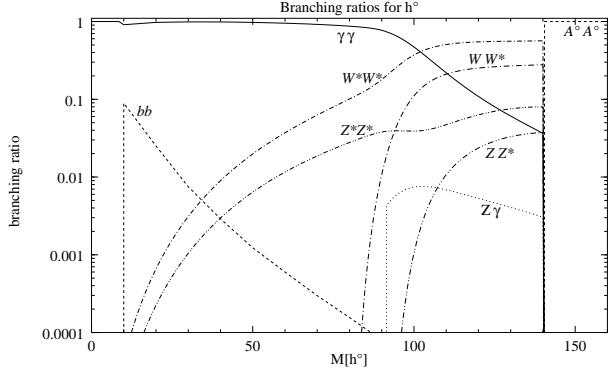,width=8cm}
    \caption{Branching ratios of $ h^0 $ at $m_{A^0}=70\ GeV$, $m_{H^+}=140\ GeV$, $m_{H^0}=300\ GeV$ and $\delta= 0.1$ in potential $A$. }
    \label{fig:h0A1}
  \end{center}
\end{figure}

\begin{figure}[htbp]
  \begin{center}
    \epsfig{file=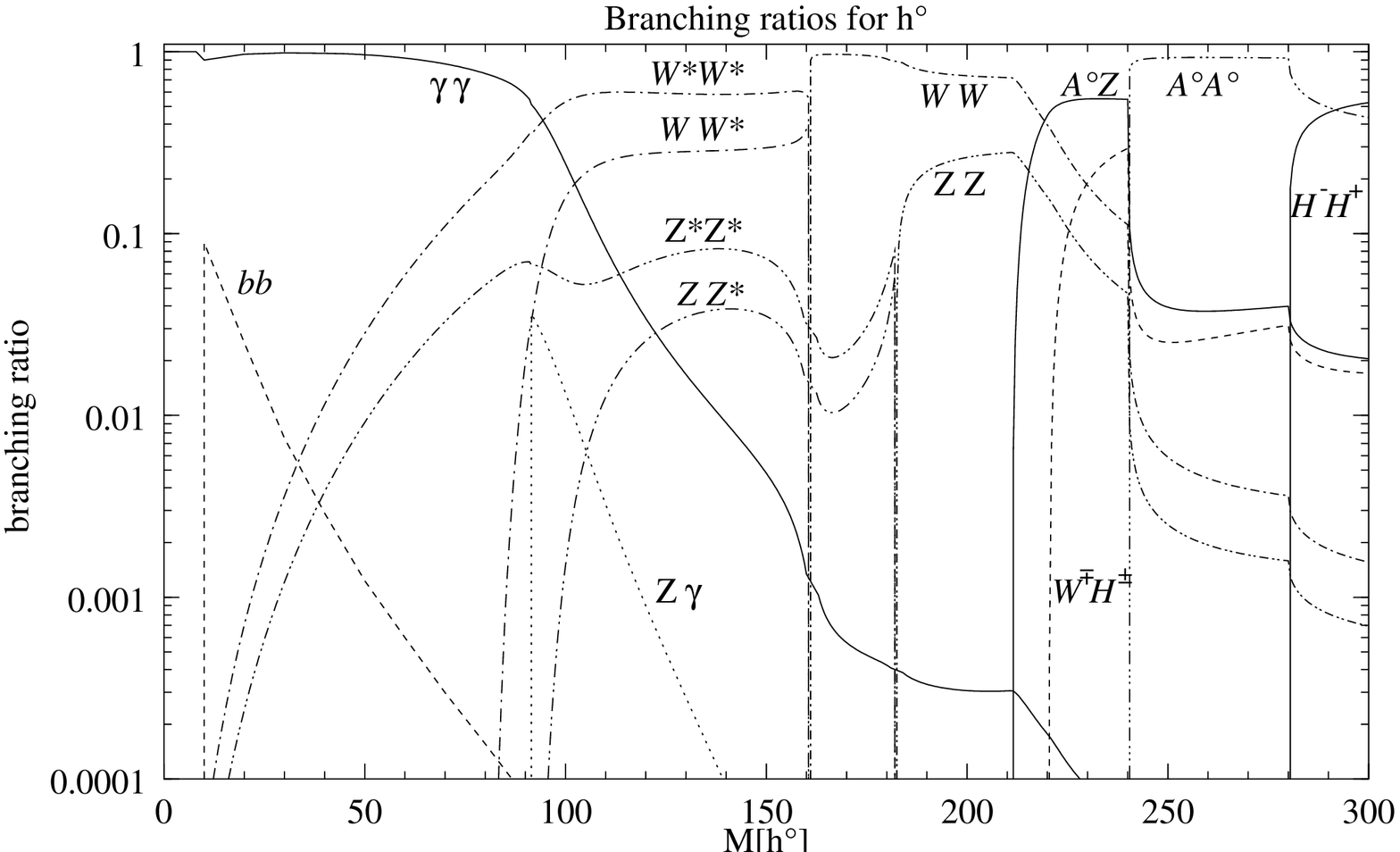,width=8cm}
    \caption{Branching ratios of $h^0$ at $m_{A^0}=120\ GeV$, $m_{H^+}=140\ GeV$, $m_{H^0}=300\ GeV$ and $\delta= 0.2$ in potential $A$. }
    \label{fig:h0A3}
  \end{center}
\end{figure}

In potential $A$ the upper bound for the mass of the lightest scalar
Higgs boson is approximately the $W$ mass in the tiny $\delta$ region.
Thus $h^0$ has only two possible decay modes. Either it decays into
$A^0 A^0$, if the mass of the lightest scalar is twice as large as the
mass of the pseudo-scalar Higgs boson, or it decays into two
photons.\footnote{The third possible decay, $h^0\rightarrow H^+H^-$ is
  already ruled out by the experimental lower limit on the mass of the
  charged Higgs boson (cf. section \ref{exlim}).} In the small
$\delta$ region the growth of the upper mass limit for $m_{h^0}$ gives
rise to more decay modes, as can be seen in fig.~\ref{fig:h0A1}. For
small $h^0$ masses the situation is the same as in the tiny $\delta$
region.  Depending on the mass of the pseudo-scalar, the dominant
decay is again either $h^0 \rightarrow A^0 A^0$ or
$h^0\rightarrow\gamma\gamma$. As soon as $m_{h^0} > m_W$, decays via
virtual vector bosons overtake the decay to $\gamma\gamma$ and give
rise to a fermionic signature of $h^0$.  Of course the value of
$m_{h^0}$, for which the branching ratio of $h^0\rightarrow W^*W^*$
becomes bigger than 50\% depends on $\delta$.  At the lower end of the
small $\delta$ region this happens approximately at $m_{h^0}=110\ 
GeV$, whereas at the upper end it is close to the $W$ mass. At first,
in the large $\delta$ region the branching ratio does not change much.
Of course the upper bound for $m_{h^0}$ looses importance and all
decays become kinematically allowed, as can be seen in
fig.~\ref{fig:h0A3}.  As $\delta$ increases, the decay $h^0\rightarrow
b \bar{b}$ becomes more and more significant for small masses of
$m_{h^0}$. If e.g. $m_{h^0}=20\ GeV$ we get a branching ratio for
$h^0\rightarrow b \bar{b}$ of the order of $30\%$ at $\delta=0.5$ and of
$75\%$ at $\delta=1.0$. This reflects the already mentioned
$\tan^2\delta$ suppression of this decay mode.

\begin{figure}[htbp]
  \begin{center}
    \epsfig{file=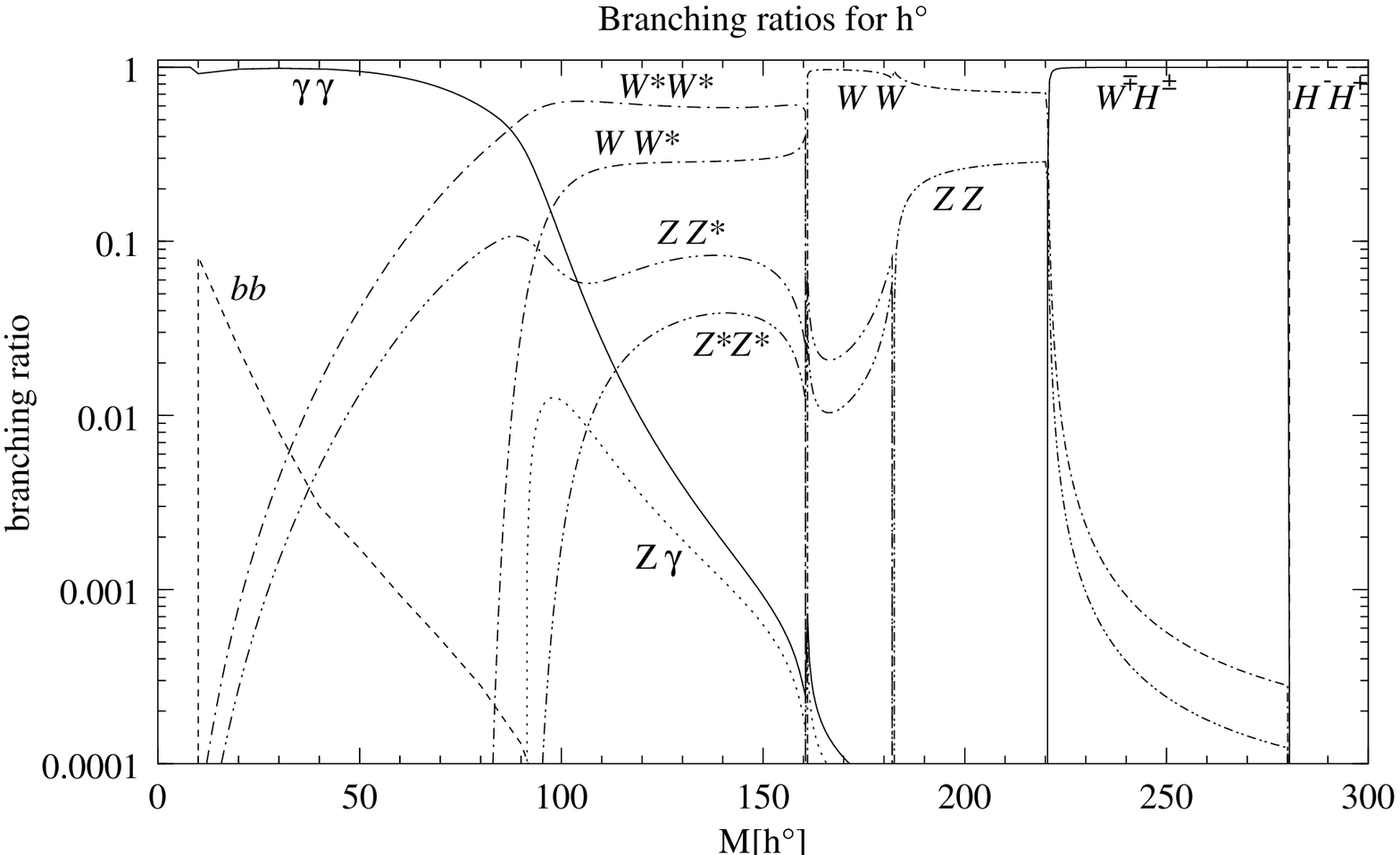,width=8cm}
    \caption{Branching ratios of $h^0$ at $m_{A^0}=m_{h^0}$, $m_{H^+}=140\ GeV$, $m_{H^0}=300\ GeV$ and $\delta= 0.01$ in potential $B$. }
    \label{fig:h0B1}
  \end{center}
\end{figure}

\begin{figure}[htbp]
  \begin{center}
    \epsfig{file=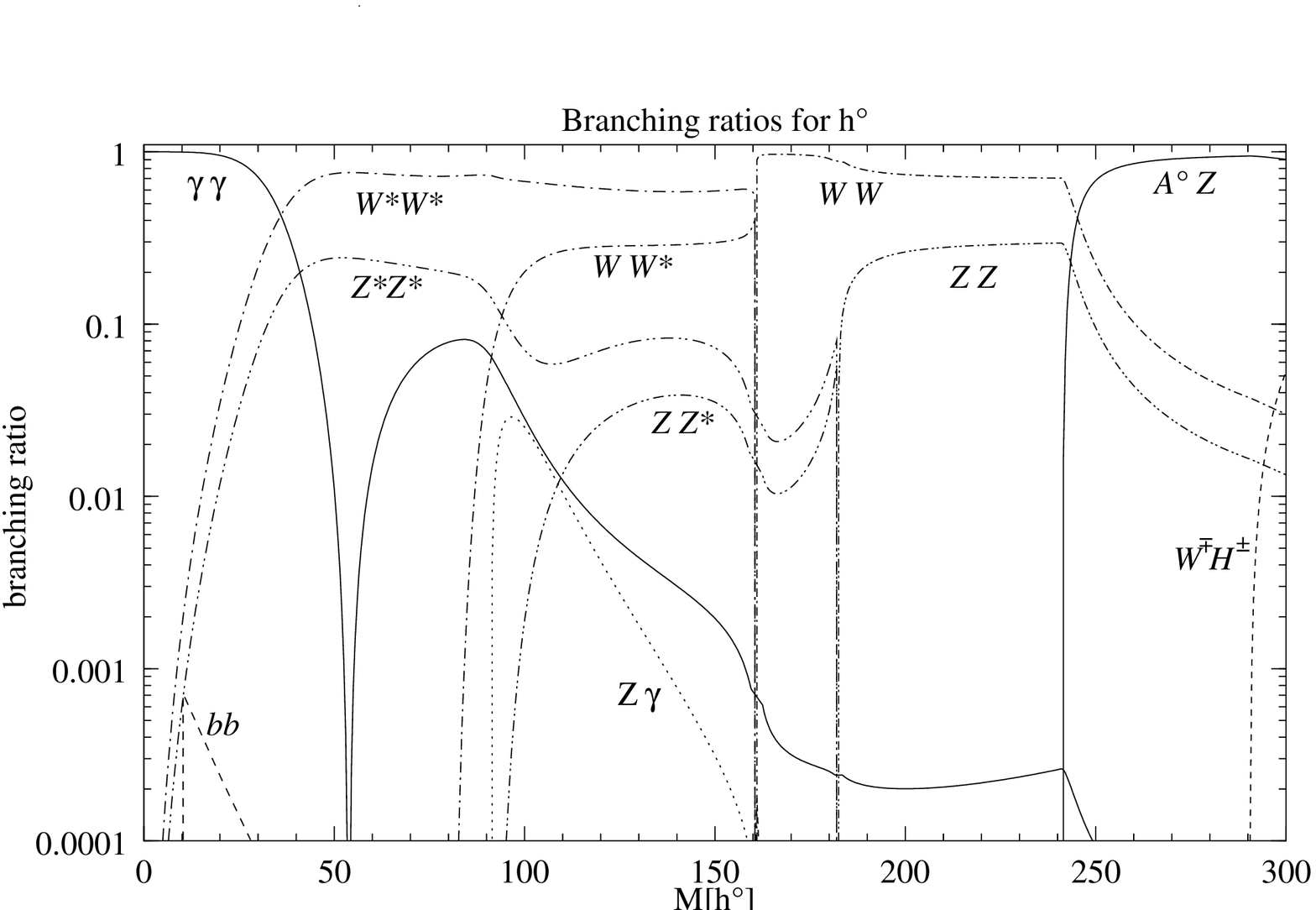,width=8cm}
    \caption{Branching ratios of $h^0$ at $m_{A^0}=150\ GeV$, $m_{H^+}=210\ GeV$, $m_{H^0}=300\ GeV$ and $\delta= 0.1$ in potential $B$. }
    \label{fig:h0B2}
  \end{center}
\end{figure}

In potential $B$ the masses of $h^0$ and $A^0$ are almost degenerated
in the tiny $\delta$ region. Thus for small masses ($<m_W$) $h^0$
decays mainly into two photons. On the other hand, no upper bound on
$m_{h^0}$ exists in potential $B$. As a consequence a heavy $h^0$ can
also decay via virtual vector bosons into fermions in the tiny
$\delta$ region (cf. fig.~\ref{fig:h0B1}).  In the small $\delta$
region the branching ratio strongly depends on the parameters $m_A$
and $m_{H^+}$. It can either resemble the plot for potential $A$ (see
fig.~\ref{fig:h0A1}), or, due to strong cancellation between the
$H^+$- and the $W$-loops in the $h^0\rightarrow\gamma\gamma$ decay, it
can be as shown in fig.~\ref{fig:h0B2}. In this figure we see that
$h^0\rightarrow\gamma\gamma$ only dominates until $m_{h^0}\approx 30\ 
GeV$. Then, decays via virtual vector bosons are the major decays of
$h^0$.  Note that $h^0\rightarrow b \bar{b}$ is suppressed in a
similar way to $h^0\rightarrow\gamma\gamma$, because both decays
depend on the same couplings of $h^0$ to the vector bosons and to the
scalars. In the large $\delta$ region this behaviour is almost the
same.  Of course, as in potential $A$, for some value of $\delta$ the
decay $h^0\rightarrow b \bar{b}$ will dominate over
$h^0\rightarrow\gamma\gamma$ for small values of $m_{h^0}$.

\begin{figure}[htbp]
  \begin{center}
    \epsfig{file=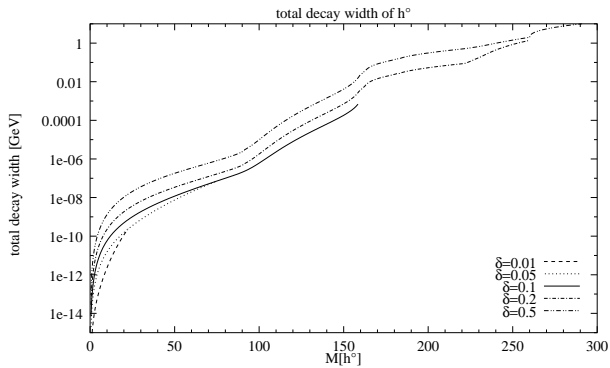,width=8cm}
    \caption{Total decay width of $h^0$ with $m_{A^0}=130\ GeV$,
      $m_{H^+}=150\ GeV$, $m_{H^0}=300\ GeV$ for different values of
      $\delta$ potential $A$. }
    \label{fig:h0A2}
  \end{center}
\end{figure}

Finally we show the total decay width of $h^0$ as function of
$m_{h^0}$ for different values of $\delta$ in fig.~\ref{fig:h0A2}. As
expected, the total decay width grows with $m_{h^0}$ and $\delta$.  We
do not show the total decay width for potential $B$ because the
overall behaviour is the same as for potential $A$.


\section{Constraints on the models}\label{exlim}

\begin{figure}[htbp]
  \begin{center}
    \epsfig{file=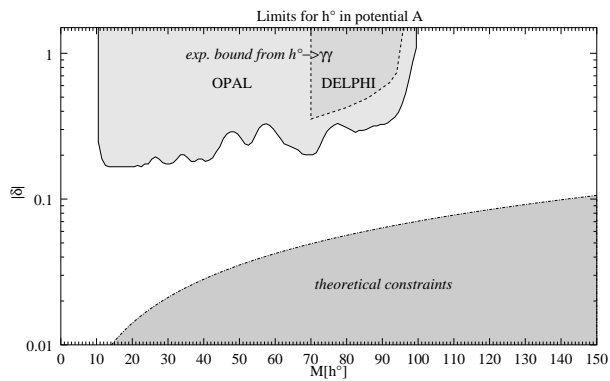,width=8cm}
    \caption{Bounds in the $m_{h^0}$-$\delta$ plane for
      potential $A$.}
    \label{fig:LimhA}
  \end{center}
\end{figure}

\begin{figure}[htbp]
  \begin{center}
    \epsfig{file=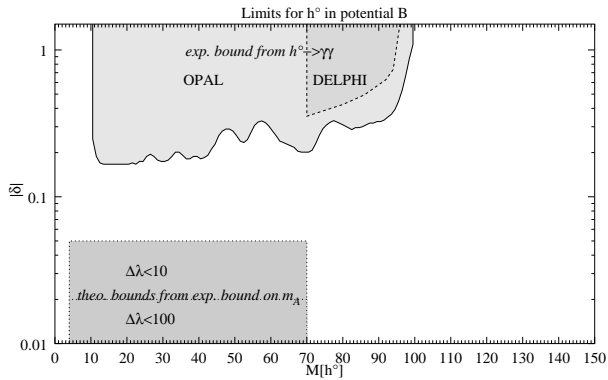,width=8cm}
    \caption{Bounds in the $m_{h^0}$-$\delta$ plane for
      potential $B$.}
    \label{fig:LimhB}
  \end{center}
\end{figure}

In this section we use the available experimental data and the bounds
derived in section \ref{masslim} to constrain the models.

Most production modes of the pseudo-scalar Higgs boson at LEP are
suppressed in the fermiophobic limit. An exception is the associated
production $Z^*\rightarrow h^0 A^0$ when kinematically allowed. The more
$\delta$ tends to zero the larger becomes the cross section for this
production mode. However, the obtained limit for $m_A$ is not independent of
the mass of the lightest scalar Higgs boson. This production mechanism
has recently been measured by the DELPHI coll. \cite{hex2}, where more
detailed results
can be found. For this associated production we roughly summarize their result in the following
inequation:
\begin{equation}
  \sqrt{m_{h^0}^2 + m_A^2} \ge 80\ GeV
\end{equation}

For the lightest scalar Higgs boson mass the most stringent bounds can
be derived from the experimental measurement of massive di-photon
resonances. The most recent data have been published in refs.
\cite{hex1,hex2}. We have used this data to exclude some regions in
the $m_{h^0}$-$\delta$ plane. We have plotted the results in
fig.~\ref{fig:LimhA} for potential $A$ and in fig.~\ref{fig:LimhB} for
potential $B$. 
Moreover we have inserted the
theoretical constraints shown in fig.~\ref{fig:Mhlimit1A}. In
fig.~\ref{fig:LimhA} (potential $A$) this can be seen as the lower
limit on $\delta$ for a given $h^0$ mass. For potential $B$ the
experimental bound on $m_A$ can be used to derive a lower limit on
$\delta$ for a given $m_{h^0}$. In fig.~\ref{fig:LimhB} we have
plotted this area for different values of
$\Delta\lambda$.\footnote{c.f.  section \ref{masslim}.}

\section{Conclusion and outlook}

We have shown the branching ratios for the lightest $CP$-even scalar
Higgs particles of fermiophobic 2HDM´s as a function of the Higgs
masses and $\delta$. We have shown that the two different scalar
sectors, potential $A$ and $B$, give rise to different signatures for
some regions of the parameter space. Most of the mass bounds based on
a general 2HDM or on the MSSM do not apply in the fermiophobic
case. We have used the available experimental data and tree-level
unitarity bounds to constrain the models. It turns out, that there is
still a wide region of this parameter space not yet excluded by
experimental data and still accessible at the LEP collider. So, one
should keep an open mind for surprises in the Higgs sector.

\section*{Acknowledgments}

We like to thank our experimental colleagues at LIP for the
inspiring discussions. L.B. is partially supported by JNICT contract No.
BPD.16372.


\end{document}